\documentclass{mem}
\usepackage{natbib}
\usepackage{txfonts}
\usepackage{balance}
\usepackage{graphicx}
\usepackage[a4paper]{hyperref}
\idline{75}{282}
\begin{document}

\title{Axisymmetric mass models of S0 and spiral galaxies with boxy
bulges: mass-to-light ratios, dark matter and bars}

   \subtitle{}

\author{Michael J.~Williams, Martin Bureau \and Michele Capppellari}

\institute{
Sub-department of Astrophysics, Denys Wilkinson Building, University of Oxford, Oxford OX1 3RH, United Kingdom,
\email{williams@astro.ox.ac.uk}
}

\authorrunning{Williams, Bureau \& Cappellari}

\titlerunning{Axisymmetric mass models of S0s and spirals}

\abstract{We examine a sample of 30 edge-on spiral and S0 galaxies that
have boxy and peanut-shaped bulges. We compute model stellar kinematics by
solving the Jeans equations for axisymmetric mass distributions derived
from $K$-band images. These simple models have only one free parameter:
the dynamical mass-to-light ratio, which we assume is independent of
radius. Given the simplicity of the modelling procedure, the model
second velocity moments are strikingly good fits to the observed stellar
kinematics within the extent of our kinematic data, which typically
reach $\approx$ 0.5--1\,$R_{25}$ (where $R_{25}$ is the optical radius),
or equivalently $\approx$ 2--3\,$R_\mathrm{e}$ (where $R_\mathrm{e}$ is
the effective or half-light radius). We therefore find no evidence for a
dominant dark matter component within the optical disk of spiral
galaxies. This is equally true of the S0s in our sample, which
significantly extends previous observational constraints on dark matter
in these galaxies. The predicted kinematics do deviate slightly but
systematically from the observations in the bulge region of most
galaxies, but we argue that this is consistent with the claim that boxy
and peanut-shaped bulges are bars viewed edge-on. 
}
\maketitle{}

\section{Introduction}

Previous studies indicate with some certainty that dark matter does not
make a significant contribution to the total mass within the optical radius,
$R_{25}$, of disk galaxies
\citep[e.g.][]{Palunas:2000,Bell:2001,Kassin:2006}. Similar studies of
ellipticals find little evidence for significant dark matter within the
effective radius, $R_\mathrm{e}$
\citep[e.g.][]{Gerhard:2001,Rusin:2003,Romanowsky:2003,Cappellari:2006,Thomas:2007}.
However, degeneracies in the modelling mean that the dark matter content
of ellipticals is more uncertain. Moreover, $R_\mathrm{e} \ll R_{25}$.

Non-axisymmetric galaxies make it possible to lift degeneracies in the
contributions from luminous and dark matter by ascribing non-circular
motions to luminous matter only
\citep[e.g.][]{Englmaier:1999,Weiner:2001}. Fast observed pattern speeds
also constrain dark matter in barred disks
\citep[e.g.][]{Debattista:2000,Aguerri:2003,Gerssen:2003}.

In this study we use the sample of thirty S0 and spiral galaxies with
a boxy or peanut-shaped bulge of \cite{Bureau:1999} to constrain
their dark matter content. From previous numerical and
observational work, such galaxies are thought to be barred disk
galaxies viewed edge-on
\citep[e.g.][]{Kuijken:1995,Bureau:1999,Chung:2004,Mendez-Abreu:2008}.

\section{Method and assumptions}

We modelled the stellar kinematics of the sample galaxies by making two
key assumptions: that dark matter is an insignificant dynamical
component and that these galaxies are axisymmetric (so are unbarred). In
this way, we were able to constrain the dark matter content while
simultaneously seeking independent confirmation that boxy bulges are
bars viewed edge-on.

We first parametrized the apparent light distributions of the galaxies
using $K$-band images \citep{Bureau:2006} and the multi-Gaussian
expansion method \citep{Emsellem:1994}. We deprojected these light
distributions by assuming that the galaxies are axisymmetric. We then
converted the intrinsic light distributions to intrinsic mass
distributions by assuming constant mass-to-light ratios. This is
equivalent to assuming either that there is no significant dark matter,
or that the dark matter distribution follows the visible matter in shape
and scale (which would be inconsistent with CDM). Finally, we computed
predictions of the stellar second velocity moments integrated along the
line-of-sight by solving the Jeans equations under the assumption of
constant anisotropy in the meridional plane (see
\citealt{Cappellari:2008} for details). 

\begin{figure*}[t]
\label{fig:results}
\centering
\includegraphics{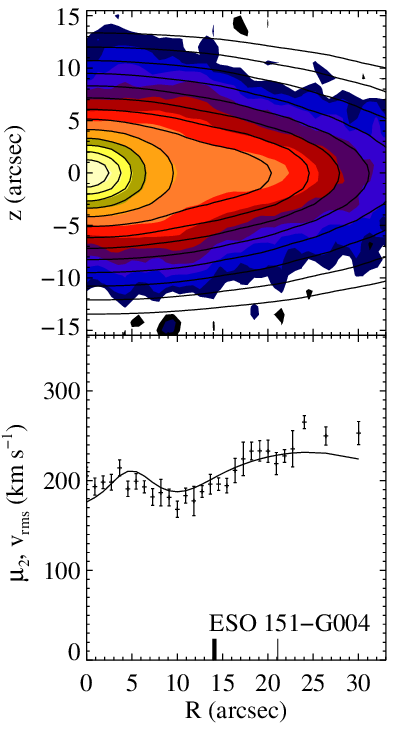}
\includegraphics{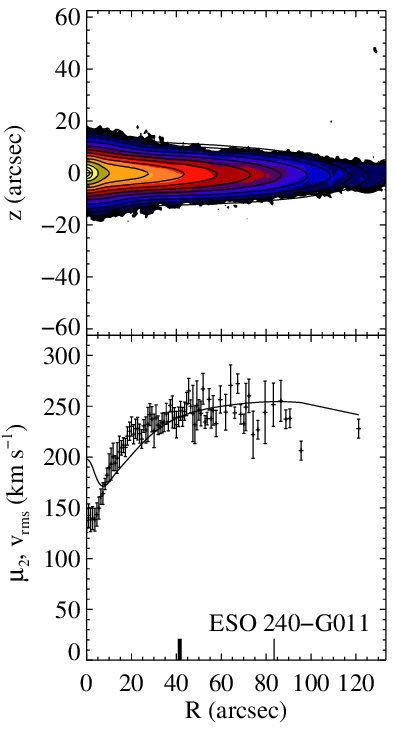}
\includegraphics{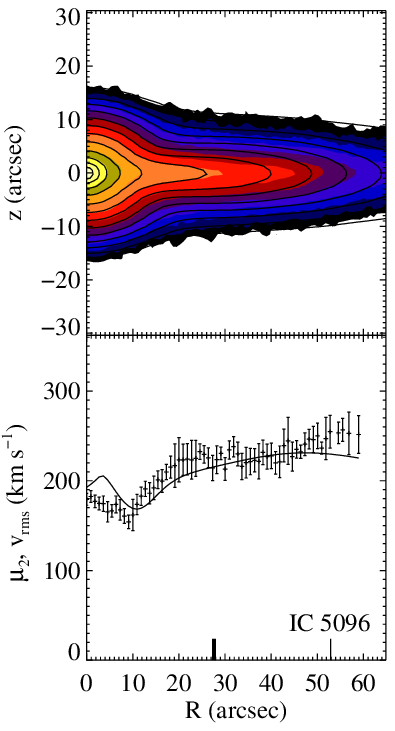}
\caption{Stellar dynamical models of 9 of the 30 galaxies in our sample.
The top pane of each figure shows the $K$-band
image (filled contours) and the MGE fit from which the mass model was
generated (lines). The bottom pane shows the observed stellar
$v_\mathrm{rms} \equiv (v_\mathrm{obs}^2 + \sigma_\mathrm{obs}^2)^{1/2}$
(points) and the model second velocity moment (line). The thick notches
on the R-axes are at 1\,$R_\mathrm{e}$ and the thin notches at
0.5\,$R_{25}$. The models shown assume isotropy in the meridional plane,
but we also experimented with fixed, non-zero anisotropies (see
Section~\ref{sec:results}).} \end{figure*}

\section{Results}
\label{sec:results}

Results for 9 of the 30 galaxies are shown in Figure~1 (see
\citealt{Williams:2008} for figures for the complete sample). Bearing in
mind that our modelling procedure contains just one free parameter, the
mass-to-light ratio (which simply scales the models up and down), the
model second moments are strikingly accurate representations of the
kinematic data (themselves from \citealt{Chung:2004}). The models shown
assume isotropy ($\sigma_z = \sigma_R$, $\sigma_{Rz} = 0$). We
experimented with anisotropic models with $\sigma_z < \sigma_R$ and
found that, outside the bulge, it did not significantly affect the
models. There is a slight but systematic deviation of the models within
the central 10 arcsec or so, a region which roughly corresponds to the
boxy bulges (and thus the non-axisymmetric barred regions). 

\section{Discussion}

Because our models are such good fits to the data, they provide no
evidence for a dominant dark component within 0.5--1$ R_{25}$ or 2--4
$R_\mathrm{e}$ in a sample of thirty S0 and spiral galaxies. This is
consistent with previous studies of spiral galaxies. It also extends the
radius out to which observational constraints may be placed on the dark
matter content of S0s. In fact, since S0s have been demonstrated to be
observationally equivalent to fast-rotating ellipticals
\citep{Emsellem:2007,Cappellari:2007}, this statement may also be
tentatively applied to fast-rotating ellipticals.

The $K$-band dynamical mass-to-light ratios that we measure are
typically in the range 1--2\,$M_\odot/L_{\odot,K}$. This is consistent with
those of the SAURON sample \citep{Cappellari:2006}, but the mass-to-light
ratios predicted by the single stellar population models of
\cite{Maraston:2005} with a Kroupa initial mass function do not exceed
1.4\,$M_\odot/L_{\odot,K}$. This may thus hint at the presence of 
dynamically subdominant dark haloes. To place constraints on
the amount, we have constructed Jeans models which explicitly include
dark halos \citep{Williams:2008}

We are limited by the small size of our non-boxy control sample, but the
slight but systematic deviation of the models within the bulge regions
is consistent with the claim that boxy bulges are bars viewed edge-on.
The sense of the discrepancy (model $>$ data) is also consistent with
the construction of the sample, which is dominated by bars viewed
side-on (end-on bars appear spheroidal so were not included in this boxy
sample). The radial extent of this discrepancy is smaller than the bars,
which are probably about twice the length of the boxy bulge
\citep{Athanassoula:2005}, but simple models have shown that the axial
ratio of the $x_1$ orbits supporting the bar decreases rapidly toward
unity from its peak at around 20 percent of the length of the bar
\citep[see, e.g., Figure~2~(a) of][]{Bureau:1999a}.

We will follow up the present work by using the derived dynamical
mass-to-light ratios in two ways. Firstly, we will compare them to those
inferred from stellar population syntheis models based on absorption
line strength indices, providing further constraints on the dark matter
content and initial mass function of the galaxies. Secondly we will
compute total dynamical masses and study the Tully-Fisher relation of
these galaxies, which should reveal information about the possible
evolutionary link between spirals and S0s.

\begin{figure*}
\centering
\label{fig:results}
\includegraphics{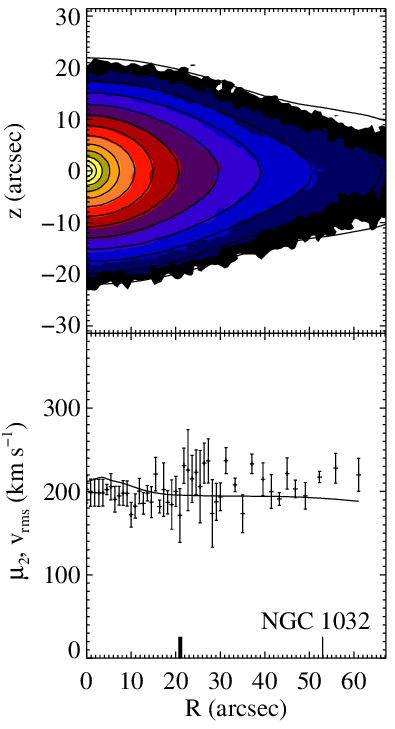}
\includegraphics{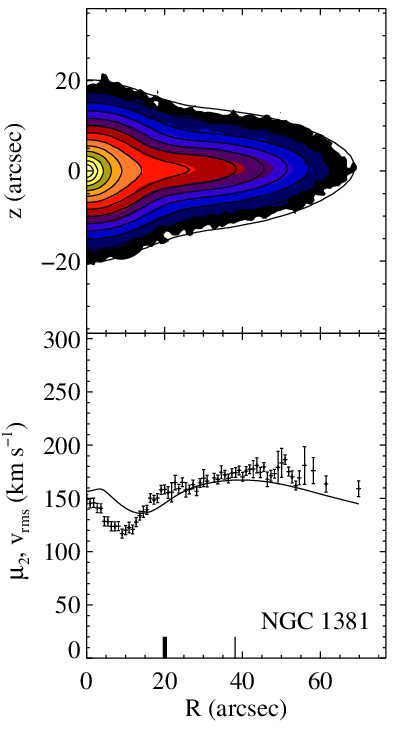}
\includegraphics{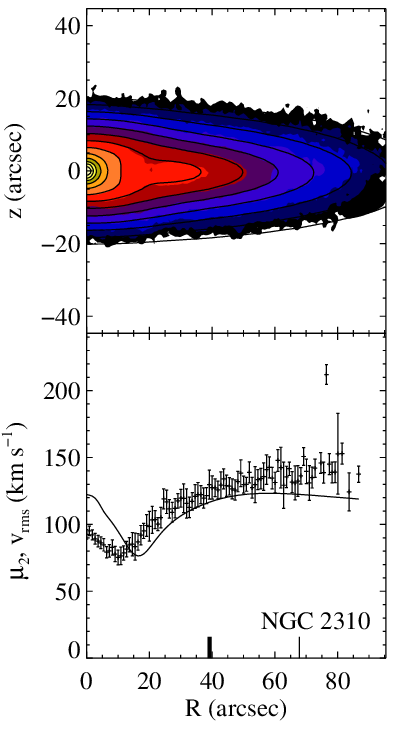}
\includegraphics{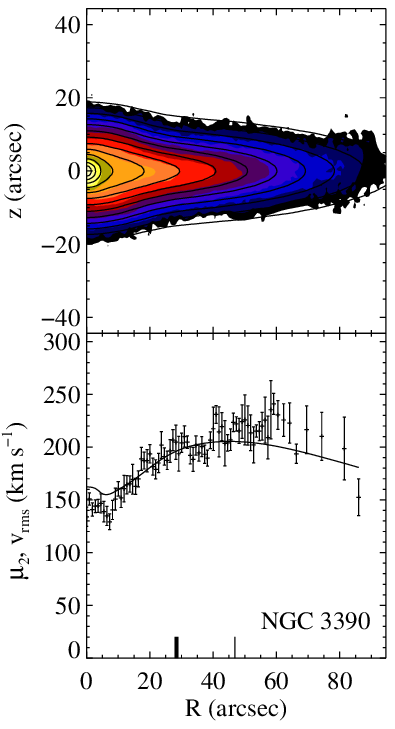}
\includegraphics{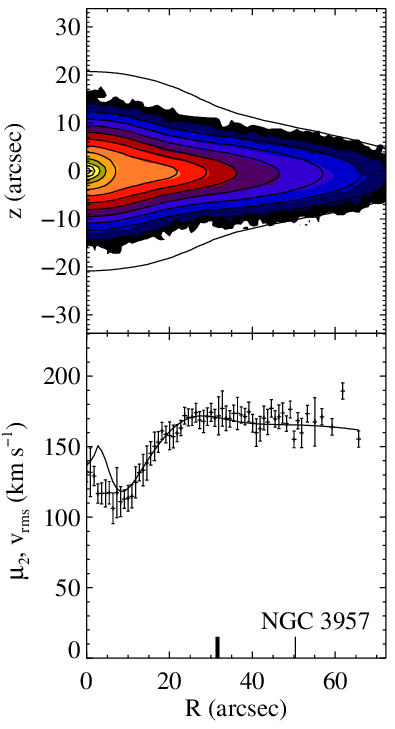}
\includegraphics{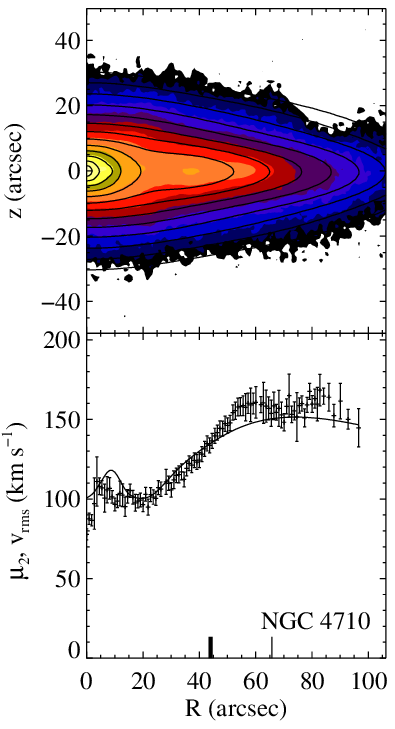}
\addtocounter{figure}{-1}
\caption{ --- continued}
\end{figure*}


\bibliographystyle{aa}

\end{document}